\documentclass[twocolumn,showpacs,preprintnumbers,amsmath,amssymb,groupedaddress,superscriptaddress]{revtex4}

\usepackage[dvips]{graphicx}
\usepackage[dvips]{color}
\DeclareGraphicsExtensions{.eps} 

\usepackage{dcolumn}
\usepackage{bm}
\newcommand{\ignore}[1]{}

\begin{document}
\def\e{\mathcal{E}}

\title{Transition linewidth of cross-correlations in random intensity fluctuations in EIT}
\author{Lei Feng}
\affiliation{Department of Physics, State Key Laboratory of Surface Physics,  Laboratory of Advanced Materials and Key Laboratory of Micro and Nano Photonic Structures (Ministry of Education), Fudan University, Shanghai 200433, China}
\author{Pengxiong Li}
\affiliation{Department of Physics, State Key Laboratory of Surface Physics,  Laboratory of Advanced Materials and Key Laboratory of Micro and Nano Photonic Structures (Ministry of Education), Fudan University, Shanghai 200433, China}

\author{Mengzhen Zhang}
\affiliation{Department of Physics, State Key Laboratory of Surface Physics,  Laboratory of Advanced Materials and Key Laboratory of Micro and Nano Photonic Structures (Ministry of Education), Fudan University, Shanghai 200433, China}

\author{Tun Wang}
\affiliation{Department of Physics, University of Connecticut, Storrs, Connecticut 06269, USA}
\affiliation{Institute of Care-life, Chengdu 610041, China}
\author{Yanhong Xiao}
\affiliation{Department of Physics, State Key Laboratory of Surface Physics,  Laboratory of Advanced Materials and Key Laboratory of Micro and Nano Photonic Structures (Ministry of Education), Fudan University, Shanghai 200433, China}
\date{\today}

\begin{abstract}

It is known that cross-correlation between the random intensity fluctuations of two
lasers forming electromagnetically-induced transparency (EIT) exhibits
a transition from correlation to anticorrelation. We study the linewidth behavior of this transition
and have found the linewidth is below the (effective) coherence lifetime limit and is only limited by competing noises.
We established a numerical model which reveals the linewidth dependence on laser linewidth and laser power.
Our experiments using lasers with different linewidth showed results in qualitative agreement with the model.
This result is useful for quantum optics using EIT, and may also have applications in spectroscopy and precision measurements.

\end{abstract}


%

\maketitle
\section{\label{intro} Introduction}
Manipulating photon statistics and laser noise through coherent atom-light interaction has recently become
a topic of great interest for quantum optics, spectroscopy and quantum information processing.
Electromagnetically induced transparency (EIT)~\cite{EITreview} or coherent population trapping (CPT)~\cite{CPTreview}
is one of the most widely used processes for such purposes. Interesting findings on both the quantum fluctuations and the classical noises in
EIT/CPT fields have been reported. For instance, a CPT system can be used to generate entangled bright photon pairs~\cite{CPTsqueezing1}
and spin squeezing simultaneously~\cite{CPTsqueezing2}; EIT is opaque to quantum fluctuations
in laser fields~\cite{OpacityEIT}; EIT-like features exist in the noise spectrum of lasers ~\cite{Godone,Lezama,XiaoPRA}
even when far off EIT resonance.

In this paper, we study the linewidth of the transition from correlation to anticorrelation in the random intensity fluctuations of
the two EIT fields.  Random phase noise in a laser can be converted to random intensity fluctuations after resonant laser-atom interactions.
Previously, it was found that such random intensity fluctuations in the two EIT fields exhibit a transition from correlation to anticorrelation when
one detunes from EIT resonance, and the transition had a linewidth narrower than that of EIT~\cite{Sautenkov,XiaoPRA,Ariunbold,Felinto}, and seemed not to be power broadened. However, it remains unclear what determines
the cross-correlation resonance linewidth, and whether this linewidth can be even narrower than the non-power-broadened EIT width, i.e., the coherence lifetime limited width.

Here, we report the finding that the linewidth is below the coherence lifetime limit and depends on the laser linewidth and laser power. In principle the linewidth can approach zero for infinitely small laser
width if other noise sources are neglected. We established a numerical model (section \ref{Theory}) which predicts linewidth behaviors in
good qualitative agreement with our experimental results (section \ref{Experiment}). Although our current study is on classical noise,
the result is meaningful for quantum noise experiments since laser phase noise is inevitable and laser phase noise to intensity noise conversion (PN-IN) ~\cite{Camparo} may overwhelm quantum correlations. Meanwhile, this study can contribute to the interesting field
of noise spectroscopy~\cite{Noisespectroscopy1,Noisespectroscopy2}, and can also be useful for precision measurement purpose since such PN-IN process degrades performance of atomic frequency standards~\cite{Godone}.

\section{\label{Theory} Theory}

\begin{figure}[t]
\includegraphics[clip,width=0.7\linewidth]{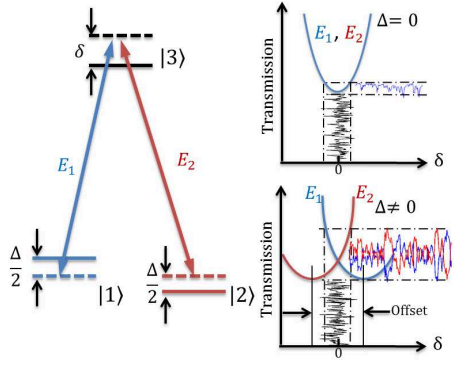}
\caption{(Color online) Intuitive picture for the sharp transition from correlation to anticorrelation in the intensity fluctuations of the two EIT fields $E_1$ and $E_2$. $\delta$ is the averaged one-photon detuning, and $\Delta$ is the two-photon detuning, generated by shifting the energy levels with a magnetic field via Zeeman shift. \it{(See text for details).}}
\label{level.fig}
\end{figure}

The physics picture for the sharp transition from correlation to anticorrelation was provided in~\cite{ourFM} for
frequency-modulated lasers.  It is still applicable here since the random laser noise
can be viewed as the sum of frequency modulation at all frequencies~\cite{noise-OL}.
%
We consider a typical CPT configuration (see Fig.~\ref{level.fig}) where
the pump $E_1$ and probe $E_2$ have equal Rabi frequencies and identical phase noises originated from the same laser source.
They couple the excited state $|1\rangle$
to two ground states $|2\rangle$, $|3\rangle$ respectively. Although the physics is valid for
a generic $\Lambda$ system, we consider the Zeeman EIT for simplicity. When the two-photon detuning
$\Delta=0$, $E_1$ and $E_2$ have the same one-photon detuning, and hence experience parallel
PN-IN slopes leading to correlations; when $\Delta\ne0$, the sharp dispersion associated with the
ground state coherence breaks the symmetry and induces
a large offset ($\gg|\Delta|$) between the minima of the two transmission spectra~\cite{ourFM}.
This offset gives rise to opposite PN-IN slopes around $\delta=0$ resulting in
anticorrelated IN~\cite{Note}. This picture indicates that narrower laser linewidth leads
to sharper transition from correlation to anticorrelation,
as shall be verified by our theory and experiments below.

In our model, we assume the Rabi frequencies for $E_1$ and $E_2$ are $\Omega_1=\Omega_2=\Omega e^{i\phi(t)}$,
with $\phi(t)$ being the random phase noise obeying Gaussian statistics, known as
Wiener-Levy process~\cite{qnbook,Scullybook}:
\begin{equation}\label{phase}
<\dot{\phi(t)}>=0,
<\dot{\phi}(t)\dot{\phi}(t')>=2\textit{D}\delta (t-t')
\end{equation}
where $D$ is the half width of laser spectrum at half maximum (HWHM).
As in Eq.~(\ref{phase}), the phase changing rate $\dot{\phi}(t)$ does not
depend on its historical fluctuations. Dynamics of the density matrix is
determined by the stochastic optical Bloch equation~\cite{CPTbook} and the Hamiltonian is:
\begin{align}\label{Hamiltonian}
\mathbf{H}=\Omega_1 |3\rangle\langle 1| +\Omega_2 |3\rangle\langle 2|-\delta |3\rangle\langle 3|
\notag \\
+\Delta/2(|2\rangle\langle 2|-|1\rangle\langle 1|)+H.c.
\end{align}
where $\delta$ is the average one-photon detuning, $\Delta$ is the two-photon detuning
and $H.c.$ is the Hermitian conjugate. We phenomenologically introduce the excited state decay rate
$\Gamma$ with equal rates to $|1\rangle$ and $|2\rangle$, and ground state population difference (coherence)
decay rate $\gamma_1$ ($\gamma_2$). The Rabi frequency is assumed to be much smaller than $\Gamma$,
so the excited state can be eliminated and the optical coherence adiabatically
follows the ground states.

Since for typical diode lasers, the effect of phase noise on atom-light interaction dominates over that of the intrinsic
intensity noise, we neglect the latter in our model. The PN converted IN, $\delta I_1$ and $\delta I_2$
are obtained for cross-correlation calculation:
$g^{(2)}(0)=\frac{\langle\delta I_1\delta I_2\rangle}{\sqrt{\langle(\delta I_1)^{2}\rangle\langle(\delta I_2)^{2}\rangle})}$
where $\langle~\rangle$ is the time average. In the optically thin regime, intensity fluctuations are proportional to
coherence fluctuations of atoms, which gives the zero time lag cross correlation of the INs~\cite{Sautenkov}
\begin{equation}\label{cross_correlaion}
\textit{g}^{(2)} (0)=\frac{\langle\textit{Im}(\delta \rho_{31}(t)) \textit{Im}(\delta \rho_{32}(t))\rangle}{\sqrt{\langle(\textit{Im}(\delta \rho_{31}(t)))^{2}\rangle\langle(\textit{Im}(\delta \rho_{32}(t)))^{2}\rangle}}
\end{equation}
where $\rho_{ij}$ is the element of the atomic density matrix, and $\langle~\rangle$ is the ensemble average.




The average terms in Eq.~(\ref{cross_correlaion}) can be rewritten as $\langle\textit{Im}(\delta \rho_{ij}) \textit{Im}(\delta \rho_{kl})\rangle=\frac{1}{4}\lbrack \langle\rho_{ij},\rho_{lk}\rangle+\langle\rho_{ji},\rho_{kl}\rangle-\langle
\rho_{ij},\rho_{kl}\rangle-\langle\rho_{ji},\rho_{lk}\rangle\rbrack$, where $\langle A,B \rangle=\langle AB\rangle-\langle A \rangle\langle B\rangle$.
We then define a vector composed of the independent elements of the density matrix  $\mathbf{u}=(\rho_{bb},\rho_{cc},\rho_{ab},\rho_{ba},\rho_{ac},\rho_{ca},\rho_{bc},\rho_{cb})^T$
where $\rho_{aa}+\rho_{bb}+\rho_{cc}=1$ was applied. Following the perturbation
treatment in ~\cite{Stocha-book} using Eq.~(\ref{phase}),
the averaged master equation can be written as
\begin{equation}\label{stochastic1}
\partial_t \langle\mathbf{u}\rangle=[\mathbf{A}_0+2D\mathbf{B}^2]\langle\mathbf{u}\rangle+\mathbf{b}
\end{equation}
where $\mathbf{A}_0$ and $\mathbf{b}$ contain Rabi frequency, decay rates and detunings, and $\mathbf{B}$ is a constant diagonal matrix.

To compute the motion of $\langle\textit{Im}(\delta \rho_{ij}) \textit{Im}(\delta \rho_{kl})\rangle$, we define a matrix $\mathbf{U}=\mathbf{u}^{T}\mathbf{u}- \langle\mathbf{u}\rangle^{T}\langle\mathbf{u}\rangle$, and perform the same stochastic average and projection as for Eq.~(\ref{stochastic1}), and obtain
\begin{equation}\label{stochasitc2}
\partial_t \langle\mathbf{U}\rangle=[\widetilde{\mathbf{A}}_0+2D\widetilde{\mathbf{B}}^2_1]\langle\mathbf{U}\rangle
-2D(\widetilde{\mathbf{B}}_2-\widetilde{\mathbf{B}}^2_1)\langle\widetilde{\mathbf{U}}\rangle.
\end{equation}
where $\widetilde{\mathbf{A}}_0$ and $\widetilde{\mathbf{B}}_1$ are of similar origins with $\mathbf{A}_0$ and $\mathbf{B}$ in Eq.~(\ref{stochastic1}) and $\widetilde{\mathbf{B}}_2$ is a constant matrix arising in the projection from $\langle\mathbf{u}\rangle^{T}\langle\mathbf{u}\rangle$ to $\langle\widetilde{\mathbf{U}}\rangle$. Since the slowly varying envelope of $\langle\mathbf{U}\rangle$ is of our interest, the stationary solution of Eq.~(\ref{stochasitc2}) is taken to compute the cross correlation. For a two level-system, analytical solution for Eq.~(\ref{stochasitc2}) has been derived, providing a clear picture of PN to IN conversion ~\cite{ZollerPRA79,ZollerPRA88,ZollerPRA90}; however, explicit forms of the solution for a three-level system are difficult to obtain due to the much more cumbersome algebra.

\begin{figure}[t]
\includegraphics[clip,width=0.9\linewidth]{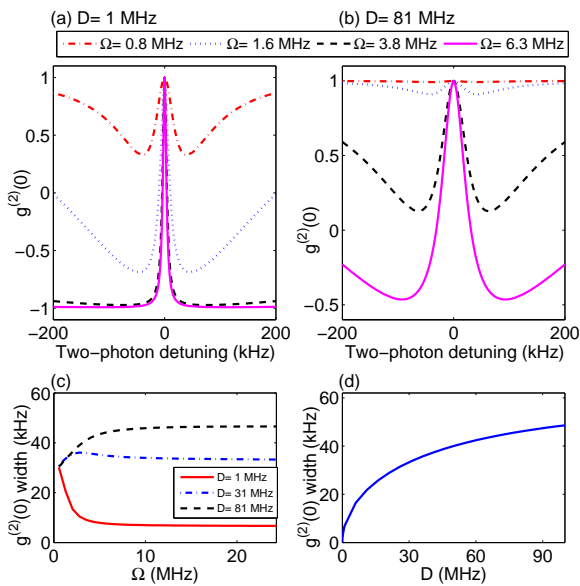}
\caption{(Color online) Numerical calculation results. (a-b) $g^{(2)}(0)$ spectra at various Rabi frequencies for laser width $D=1$ MHz and $D=81$ MHz. (c) $g^{(2)}(0)$ width vs. laser power for different laser width. (d) Saturated $g^{(2)}(0)$ width vs. laser linewidth.
Simulation parameters: ground state coherence decay rate $\gamma_2=\pi\times78$ kHz, phenomenological exited state decay rate $\Gamma=2\pi\times500$ MHz to take Doppler broadening into account. We have also performed numerical calculations where Doppler broadening was formally taken into account by integrating contributions from all velocity groups, and the excited state decay rate was set to be $2\pi \times 5.6$ MHz. This computationally heavy approach yielded $g^{(2)}(0)$ linewidth within $10 \%$ of the above result.} \label{simulation_example.fig}
\end{figure}

We have performed numerical calculation using Eq.~(\ref{stochasitc2}), and found that generally the
spectral lineshape of $g^{(2)}(0)$ vs. two-photon detuning contains a central narrow peak superimposed on a
broader dip-shaped or wing structure. It is seen that the linewidth of the wing structure increases linearly with laser power, and decreases with laser linewidth. This dependence can be explained. Anticorrelation arises from ground state coherence; since the wing structure corresponds to the disappearance of anti-correlation, its linewidth should be associated with the ground state coherence decay rate, which is equal to the optical pumping rate $\Gamma^{\prime}_{p}$ in the power broadening regime. We have $\Gamma^{\prime}_{p}=\frac{\Omega^{2}}{2(\Gamma+2D)}$, where $D$ is the laser linewidth, $\Omega$ is the total Rabi frequency and $\Gamma$ is the phenomenological excited state decay rate ($500$ MHz), and it can be seen that $\Gamma^{\prime}_{p}$ increases with the laser power and decreases with the laser linewidth. However, the narrow structure, when separated well from the wing structure, has an opposite linewidth dependence. Its linewidth decreases with the laser power and then saturate, as shown in Fig.~\ref{simulation_example.fig}(a) and (c), and increases with the laser linewidth. Similar trend has been observed and understood for the phase modulated laser case ~\cite{ourFM}, where the modulation depth plays a similar role with the laser linewidth here. The physics picture is that,
for higher laser power, the ground state coherence is stronger and hence the offset between the two EIT transmission minima is larger, making anti-correlation appear sooner, and $g^{(2)}(0)$ linewidth smaller; for larger laser linewidth, the laser frequency effectively samples
through a larger range and sees more correlation.

The linewidth of the central peak of $g^{(2)}(0)$, defined as the full width at half
of the central peak's amplitude, is then the competition result of the narrow and wing structure since the two structures are often not well separated.
This is especially the case for relatively lower laser power and larger laser linewidth, where the ground state coherence is weaker, and hence
complete anticorrelation cannot be reached. As shown in Fig.~\ref{simulation_example.fig}(c), the linewidth vs. Rabi frequency curve shows different trend for different laser linewidth. When the laser linewidth is larger, the wing width becomes smaller and the central peak becomes broader, making the central peak merge into the wing structure. Since the wing's linewidth increases with Rabi frequency, the $g^{(2)}(0)$ width increases with Rabi frequency. For smaller laser linewidth, the central peak separates better from the wing, and so the $g^{(2)}(0)$ width
behavior is dominated by that of the central peak whose width decreases with increased laser power. For intermediate laser linewidth, the
linewidth increases first and then decreases with the laser power. For all laser widths, the $g^{(2)}(0)$ width is saturated at high Rabi frequency because full anticorrelation is reached. The saturated $g^{(2)}(0)$ width increases with laser linewidth but all remains below the
lifetime limited width $2\gamma_2$ (Fig.~\ref{simulation_example.fig} (c)).

\section{\label{Experiment} Experiment}

To test the above theoretical results, experiments were carried out using a
$^{87}$Rb enriched vacuum vapor cell under CPT configuration with
ground states being Zeeman sublevels of $5^{2}S_{1/2}$, $F=2$ and exited states
being Zeeman sublevels of $5^{2}P_{1/2}$, $F^{\prime}=1$. The effective ground state
coherence decay rate in our system is determined by the transit time of the atoms
crossing the laser beam. Within a three-layer magnetic shield, the vapor cell was placed
in a solenoid which produces a homogenous magnetic field for
two-photon detuning adjustment. The laser beam was linearly polarized before entering
the cell, and its right and left circular polarized components acted as $E_1$ and $E_2$
in Fig.~\ref{level.fig} and are separately detected at the cell output by amplified photo-detectors.
ac signals were acquired at the computer from an oscilloscope and the $g^{(2)}(0)$ value was computed offline.

\begin{figure}[t]
\includegraphics[clip,width=0.9\linewidth]{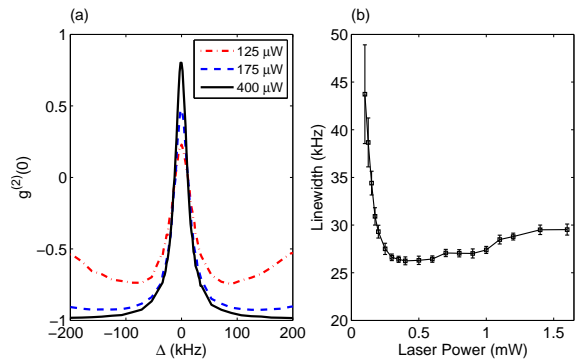}
\caption{(Color online) Experimental results using a laser with linewidth about 1 MHz. (a) Example $g^{(2)}(0)$ spectra at various laser powers.
(b) $g^{(2)}(0)$ linewidth vs. laser power.}\label{exp1.fig}
\end{figure}

We first used an external cavity diode laser with linewidth $<$1 MHz. The cell temperature
was $52^{\circ}$C. EIT FWHM vs. laser power was measured (not shown) and We extrapolated the transit time
limited width by fitting the data with polynomials and extending the fitting
result to zero power. The obtained full width is 75 kHz, close to the transit time limited width
estimated from the measured laser beam diameter of 2.8 mm. The $g^{(2)}(0)$ vs. $\Delta$ spectra at different
laser power is shown in Fig.~\ref{exp1.fig}(a). The FWHM of the central peaks of $g^{(2)}(0)$ vs. laser power are
plotted along with EIT linewidth in Fig.~\ref{exp1.fig}(b). The narrowest linewidth of $g^{(2)}(0)$ observed is about 1/3 of
the transit linewidth of EIT. At small laser power, the $g^{(2)}(0)$ spectrum can not achieve perfect anticorrelation
because of the weak ground state coherence. Besides, electronics noise from the photo-detectors makes INs less
correlated at $\Delta=0$. The trend of the $g^{(2)}(0)$ width vs. laser power is consistent with our theoretical
prediction for small laser width, except for slight broadening at large laser power. We attribute this broadening to increased influence of laser intensity noise on $g^{(2)}(0)$ at higher laser power.

\begin{figure}[t]
\includegraphics[clip,width=0.9\linewidth]{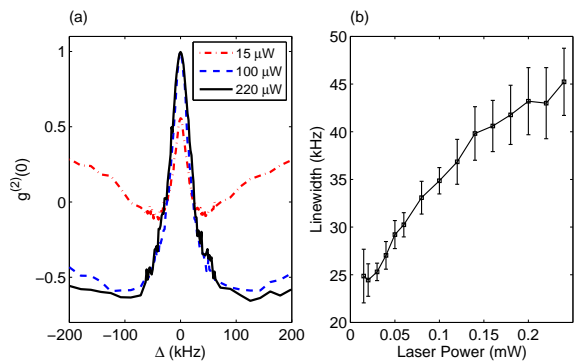}
\caption{(Color online) Experimental results using a laser with linewidth about 80 MHz. (a) Example $g^{(2)}(0)$ spectra at various laser powers.
(b) $g^{(2)}(0)$ linewidth vs. laser power.}\label{exp2.fig}
\end{figure}

In a second experiment, we used a vertical cavity surface emitting laser (VCSEL) with a linewidth of $\sim$ 80 MHz.
The vapor cell temperature was at $40^{\circ}$C
to avoid excess absorption of the relatively weak laser beam (maximal power was 250 $\mu W$) .
The $g^{(2)}(0)$ spectra [Fig.~\ref{exp2.fig}(a)] and the width [Fig.~\ref{exp2.fig}(b)] along with EIT width
are obtained at different laser power. The extrapolated transit width of EIT is 61 kHz, close to the transit width estimated from
the laser diameter of 2.5 mm.  $g^{(2)}(0)$ linewidth is also below the transit width.
The linewidth increase with laser power is consistent with the theory prediction for larger laser linewidth.

In both experiments, measured $g^{(2)}(0)$ width is larger than predicted. Several factors neglected
in our simple model contribute to the discrepancy: (a) intrinsic intensity noise from the laser in $E_1$ and $E_2$ are correlated
and hence reduces anticorrelation; (b) electronics noise in the photo detector
and data acquisition system, and eventually laser shot noise (negligible compared to other noise in our system)
are not correlated and bring the $g^{(2)}(0)$ towards zero; (c) Our system is not strictly in the
optically thin regime, and studies show that $g^{(2)}(0)$ linewidth increases with optical depth~\cite{tobepublished}.
It is due to such broadening that for the $1$ MHz laser experiment performed under $52^{\circ}$C, we observed larger $g^{(2)}(0)$ linewidth than that
in the VCSEL experiment performed under $40^{\circ}$C. ; while the model predicts that narrower linewidth always
gives narrower $g^{(2)}(0)$ linewidth as shown in Fig.~\ref{simulation_example.fig} (c).
Overall, these noise sources and complications prevent the linewidth from approaching zero as predicted in our simple model.

It is worth noting that we could only achieve 1/3 of the transit linewidth here, while in our previous work, we have obtained 1/30 of the transit linewidth using frequency modulated laser~\cite{ourFM}. We identify the following two factors that contribute to this difference. First, although the $g^{(2)}(0)$ linewidth has a positive correlation with the effective laser linewidth (including random noise and the modulated noise), one cannot compare $g^{(2)}(0)$ linewidth across these two different types of laser noises simply based on the effective linewidth, because the intrinsic random phase noise has all the frequency components and there is also a random phase between different frequency components. This tends to make the $g^{(2)}(0)$ linewidth for the random noise case broader than the modulated noise case even for the same effective laser linewidth. Second, when the laser frequency is modulated, the converted intensity fluctuation is at the same or twice the modulation frequency, which allows to average out random background noises such as residual laser intensity noise and electronics noise from the photo detectors.  For the case of random phase noise only, converted random noise and background noise cannot be separated in the detection process, and the latter tends to wash out correlation or anticorrelation. As can be seen from Fig.~\ref{exp1.fig} and Fig.~\ref{exp2.fig}, complete correlation and anticorrelation cannot be reached and therefore the $g^{(2)}(0)$ resonance is broadened.

We note that the sharp transition from correlation to anticorrelation here has an inherent connection
with the predicted sudden appearance of entanglement between collective optical modes when
Raman detuning deviates from zero ~\cite{HXM}. In ref~\cite{HXM}, a phase sensitive EIT based on
a double $\Lambda$ system is considered, where each $\Lambda$ couples two optical fields, and the two fields
coupling to the same ground state form a quantum beat and hence can be viewed as a collective field.
When the Raman detuning is zero, atoms are transparent
to light and cannot induce quantum correlations to the lasers; once
the Raman detuning deviates from zero, the sharp dispersion of the ground state coherence drastically
modifies the statistics of the fields, and entanglement occurs between any two fields from different collective modes.
In brief, the connection lies in that, in our case, atom-light interaction induces anticorrelation when exact EIT resonance condition
is broken, and in~\cite{HXM}, it introduces quantum correlation between optical fields.


\section{\label{conclusion} Conclusion}

In conclusion, we theoretically and experimentally studied the laser intensity
noise correlations generated in atom-light interactions with a phase diffused
laser in a CPT system, and found that the transition from correlation to anticorrelation
can be very sharp due to the ground state coherence, with a linewidth below the lifetime-limited width and theoretically approaching zero
at small laser width. This result is relevant to the realization of recent proposals of creating
quantum correlations between two bright optical fields and also among atomic spins using CPT, and may also be used for spectroscopy.

We thank Michael Crescimanno, Xiangming Hu, Liang Jiang, Jianming Wen, Saijun Wu and Michael Hohensee for
stimulating discussions, and Irina Novikova for suggestions in manuscript revision. This work was
funded by the NBRPC(973 Program Grants No. 2012CB921604 and No. 2011CB921604), NNSFC (Grants No. 61078013, 10904018, and J1103204), and the Research Fund for the Doctoral Program of Higher Education of China. Correspondence should be sent to yxiao@fudan.edu.cn.



\begin{thebibliography}{99}
\bibitem{Note}
The offset only exists when the ground state population difference decay rate $\gamma_1$ is nonzero.
However, the condition of anticorrelations is not the offset, but
a nonzero imaginary part of $\rho_{12}$.
\end{thebibliography}

\end{document}